\documentstyle[spie]{article} 
\input{psfig.sty}   
\def\spiffi{{\tt SPIFFI}}
\def\macao{{\tt MACAO}}

\def\sinfoni{{\tt SINFONI}}
\def\farcs{\mbox{$.\!\!^{\prime\prime}$}}

\title{A new era of spectroscopy: SINFONI, NIR integral field spectroscopy 
at the diffraction limit of an 8m telescope} 

\author{S. Mengel, F. Eisenhauer,  M. Tecza, 
N. Thatte, C. R\"ohrle, K. Bickert, and J. Schreiber\\
\skiplinehalf 
Max-Planck-Institut f\"ur extraterrestrische Physik, Postbox
1603, D--85748 Garching, Germany
\skiplinehalf 
}

\begin{document} 
  \maketitle 

%%%%%%%%%%%%%%%%%%%%%%%%%%%%%%%%%%%%%%%%%%%%%%%%%%%%%%%%%%%%% 
\begin{abstract}
\sinfoni ,  the SINgle Faint Object Near-infrared Investigation, is an
instrument for the Very Large Telescope (VLT), which will start its operation
mid 2002 and allow for the first time near infrared (NIR) integral 
field spectroscopy
at the diffraction limit of an 8-m telescope.
\sinfoni\ is the combination of two state-of-the art instruments, the integral
field spectrometer \spiffi, built by the Max-Planck-Institut f\"ur 
extraterrestrische Physik (MPE), and the adaptive optics (AO) system \macao,
built by the European Southern Observatory (ESO). It will allow a unique
type of observations by delivering simultaneously high spatial 
resolution (pixel sizes 0\farcs025 to 0\farcs25) and a moderate spectral
resolution (R$\sim$2000 to R$\sim$4500), where the higher spectral resolution 
mode will allow for software OH suppression. This opens new prospects for astronomy.
\end{abstract}

%>>>> Please include a list of keywords after the abstract 


%%%%%%%%%%%%%%%%%%%%%%%%%%%%%%%%%%%%%%%%%%%%%%%%%%%%%%%%%%%%%
\section{INTRODUCTION}
\label{sect:intro}  % \label{} allows reference to this section
A number of astrophysical questions, both extragalactic and galactic,
call for high spatial resolution imaging spectroscopy. Some
examples are the nature of high redshift objects, the centres of various types
of galaxies, quasar hosts, or, more locally, the properties of young stars
and star forming regions or the central parsec of our own Galaxy.

In June 2002, \sinfoni,\cite{Thatte98} the 
SINgle Faint Object Near-infrared Investigation, 
will start its operation at the Very Large Telescope (VLT) UT3 (KUEYEN).
\sinfoni\/ is a collaborative effort by the
Max-Planck-Institut f\"ur extraterrestrische Physik (MPE) and the
European Southern Observatory (ESO), with MPE providing the NIR
imaging spectrometer SPIFFI,\cite{Tecza98}$^,$\cite{Eisenhauer2000a}
while ESO builds the curvature sensor based
adaptive optics system MACAO.\cite{Bonaccini98} 
The laser guide star provided by the
UT3 telescope guarantees a high sky coverage for diffraction limited
observations.

\spiffi\/ 
is equipped with a 1024$^2$ HgCdTe HAWAII detector array from Rockwell.
In order to fully exploit the detector, 1024 spectra are obtained in one
integration, corresponding to a field of 32$\times$32 spatial pixels on the
sky. Each spectrum covers 1024 pixels, and the resolution can be chosen
to be R$\sim$4500 for J, H or K-band, which allows software OH suppression, 
or to be R$\sim$2000, however, covering H and K band at the
same time. The spatial pixel scale is variable (0\farcs025 to 0\farcs25) 
to allow for various observational conditions.
In order to determine \sinfoni's sensitivity, we have created an
exposure time calculator for the computation of the signal-to-noise ratio (SNR)
or the simulation of observations of template objects. 
All input parameters are
supplied by a graphical user interface. In the following chapters,
we will give a very brief technical description of \spiffi,
then describe the exposure time calculator and finally discuss some
test cases for scientific observations.

\section{SINFONI} 
The most important point for the design of a 
spectrometer at the diffraction limit of a telescope is that
the point spread function (PSF) delivered by an adaptive optics system is usually
variable on timescales of a few minutes, both in shape and in the
relative contributions of diffraction limited core versus seeing
limited halo. These variations depend in a complicated way on
parameters like seeing, magnitude of the guide star, angular
separation of guide star and object, the number of modes corrected,
the frequency of operation etc. If image deconvolution is anticipated,
information about the PSF is required. But even in the case of 
no deconvolution, the spectroscopy of extended objects requires 
the knowledge of the PSF, so that the various spatial contributions
to the spectra can be treated correctly. A long slit spectrograph
prohibits this kind of investigation, because it provides only
1-d spatial information. 
Fabry-Perot spectrometers and fourier transform interferometers deliver
the information about the PSF, but the conditions are rarely such
that the PSF stays stable over a whole wavelength range scan.
An integral field spectrometer overcomes this problem by delivering
the 2-d spatial information for the whole wavelength range in
one integration.

The NIR is best suited for AO-assisted observations, because the 
performance of AO systems increases with wavelength. 
Starting around 2.2 $\mu$m, the thermal 
background introduced by telescope and atmosphere leads to an 
increase in the noise level. A compromise of these two aspects
detemines the NIR H and K windows to be the most suited wavelenth
ranges for diffraction limited spectroscopy. \sinfoni\/
will be optimized to fulfill this task.

In short, \spiffi\/ 
is a cryogenic NIR integral field spectrometer,
based on a HAWAII focal plane array and has a variable spatial and
spectral resolution, suitable for the use with an AO system.
Some design criteria of the instrument that affect the sensitivity 
were the following:

\begin{itemize}

\item{Detector: Rockwell HAWAII focal plane array, 1024$^2$ pixels.
Readout noise 8$e^-$ per read, dark current $0.1 e^-/s$.
An upgrade to a 2048$^2$ array is foreseen.}

\item{One spectrum per column (=1024 spectra), not Nyquist
sampled in spectral domain.}

\item{Variable pixel scales to account for various observing conditions:
0\farcs025/pix for AO-assisted observations, 0\farcs1 and 0\farcs25 for
seeing limited cases.}

\item{The \spiffi\ mirror image slicer\cite{Tecza00} has a field size of
32x32 pixels (corresponding total fields between 0\farcs8$\times$0\farcs8 and 
8\farcs0$\times$8\farcs0}

\item{Spectral resolution either around 4000 for J, H or K, sufficient
for software OH suppression, or $\sim$2000, covering H and K simultaneously.
Nyquist sampling is achieved by spectral dithering in two successive 
observations.}

\item{Implementation of a ``Himmelsspinne'' (``Sky spider'') to overcome
the limitation on the length of single integrations implied by the
necessity to go to a sky position after 100s at latest. The Himmelsspinne
images a piece of sky onto the detector and allows to increase the 
integration times above 100s.}
\end{itemize}

The \spiffi\ transmission is 32\%. The transmission of the telescope and the
AO is assumed to be 80\%.

\section{THE SINFONI EXPOSURE TIME CALCULATOR}\label{etc}

The IDL-based \sinfoni\ exposure time calculator applies the basic formula

\begin{figure}[hp]
\begin{center}
\psfig{file=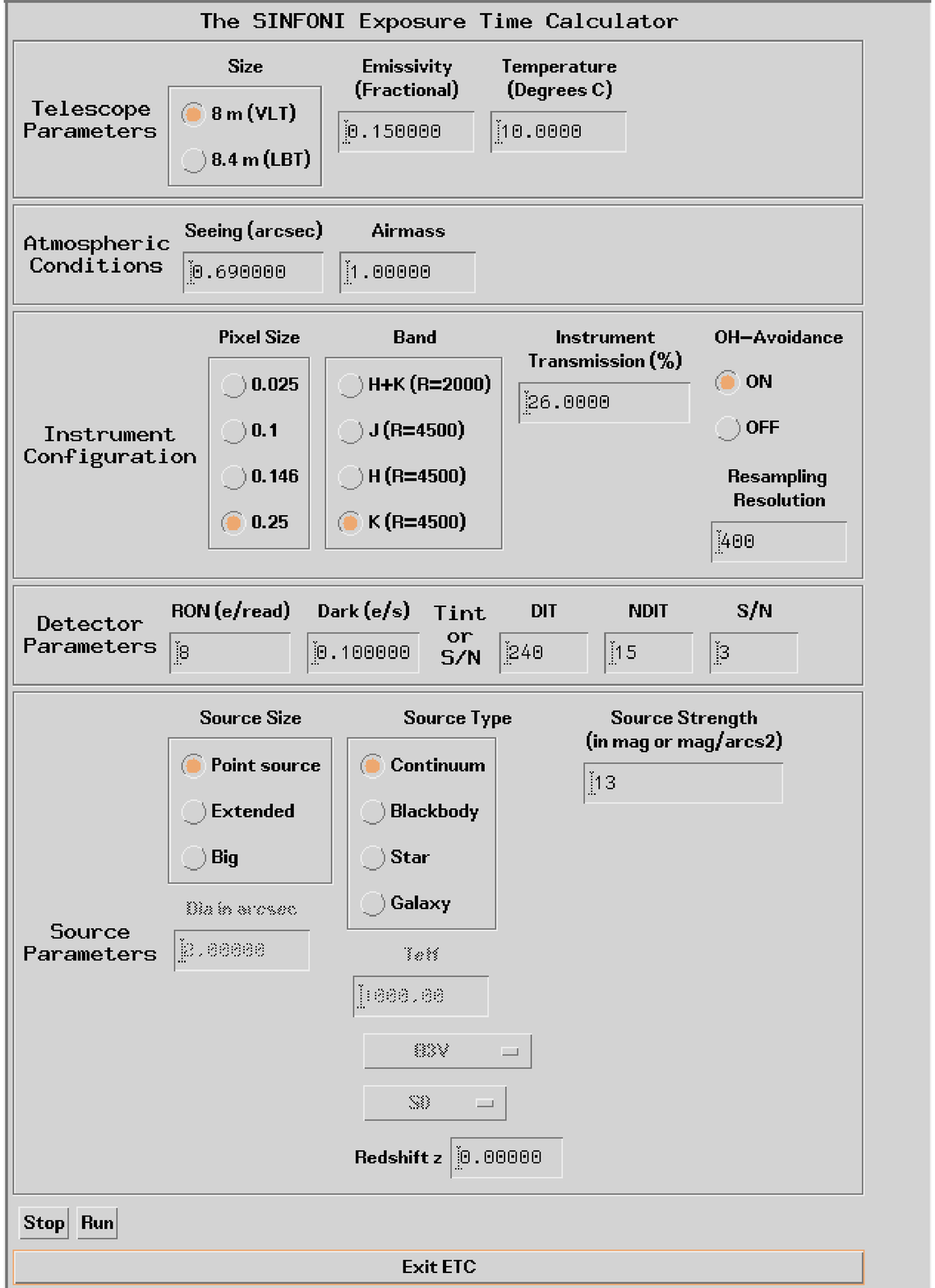,width=16cm}
\end{center}
\caption{The graphical user interface of the \sinfoni\
exposure time calculator.}
\label{etc_gui}
\end{figure}

\begin{center}
SNR  = $\langle n_s \rangle / \sqrt{N + R^2}$, N = $\langle n_s \rangle + \langle n_b 
\rangle + \langle n_d \rangle$
\end{center}

where $n_i$ denotes the total number of generated e$^-$  on a given pixel, 
the indices s, b and d indicate source, background and
dark, respectively. R is the read noise in e$^-$/read.
It is used via a graphical user interface, which is displayed in Figure \ref{etc_gui}

The parameters we take into account are:
\begin{itemize}
\item{VLT collecting area}
\item{Atmospheric transmission curve}
\item{Spectrally averaged instrument transmission}
\item{Atmospheric emission curve}
\item{Thermal background from the telescope}
\item{Detector quantum efficiency curve}
\item{Pixel size in arcsec$^2$ on the sky}
\item{Spectral resolution}
\item{Application of OH avoidance and/or spectral rebinning}
\item{Source extent (point source, extended with integrated spectrum
or extended with pixel-by-pixel analysis)}
\item{Source magnitude (in mag for point sources, mag arcsec$^{-2}$ for
extended sources), spectrum and redshift}
\item{Seeing}
\item{Strehl ratio in AO assisted observations}
\end{itemize}

Depending on what is required, two of the following parameters need to
be supplied, the third is then determined: 
\begin{itemize}
\item{Integration time}
\item{S/N} 
\item{source strength (in mag or mag arcsec$^{-2}$)}
\end{itemize}

The \spiffi\ instrument was designed to maximize the througput
wherever possible
and to minimize the background and the noise. The first goal will be
achieved by selecting high transmission/efficiency optics, and by
placing them in an evacuated environment that preserves them over a
long period of time.
The second point requires the whole spectrograph to be a cryogenic
instrument, such that the thermal background contribution of the
instrument itself is negligible compared to the telescope and the
sky. The third point was helped by the fact that the HAWAII
array we bought from Rockwell shows low read noise and dark current.
The relative impact of the read noise will be minimized by selecting long
integration times, made possible by the implementation of the sky
spider. Moreover we will optimize the electronics and the read out
modes (multiple reads). The dark current cannot be overcome, and our
estimates show that the instrument is dark limited 
at AO pixel scales and high spectral resolution in the J and the H band.

Table \ref{sensitivities} lists the sensitivities for a variety of
cases which are of interest to the general observer.
They were computed using the SINFONI exposure time calculator which
allows to set explicitly almost all of the parameters below and thus
can be used to derive the corresponding magnitudes for other cases of
interest. The limiting magnitudes given here are valid for the following 
constraints/specifications:
\begin{itemize}

\item{Operation at the VLT}
\item{S/N of 3 in 1h of integration time (15 integrations, 240s each)}
\item{Detector characteristics: 8e$^-$ read noise, 0.1e$^-$s$^{-1}$ dark 
current}
\item{Median Paranal seeing (0\farcs69), airmass 1.0}
\item{Telescope temperature 283K, emissivity 15\%}
\item{Instrumental transmission, including telescope, \macao\ and \spiffi\, but 
excluding the detector, 26\%}
\item{Wavelength dependent detector quantum efficiency as provided by Rockwell, 
average 60\%}
\item{S/N per spectral channel, resampling to R=400 for case of OH suppression}
\item{Integrated spectrum within the FWHM for the cases ``Point Source'' 
(integrated within the first Airy ring for the AO scale, within FWHM for the seeing
scales) and ``Extended'' (this means an extended source, where the signal is integrated
over an aperture supplied by the user), per spatial pixel for the last column.}
\item{Pixel size 0\farcs25 for the seeing limited, 0\farcs025 (0.5$\times$diffraction 
limit of an 8m telescope) for the AO assisted case}
\item{Strehl ratio of 0.3 for the AO case, which is achieved with a natural guide
star brighter than 14.5mag within 15'' of the object or with a laser guide
star brighter than 11th magnitude within 10'' of the object in case of K-band\cite{Thatte98}.
The strehl ratio assumed for J-band is 0.2.}
\item{Results give point source magnitudes in the case of ``Point source'' and
mag arcsec$^{-2}$ in the cases of ``Extended source'' and ``per pixel'', the average 
value for the whole band}
\end{itemize}

\begin{table}[h]
\caption{SINFONI sensitivities for various observing modes. 
1 hr integration achieving S/N=3. Point source 
magnitudes or surface brightnesses in [mag/arcsec$^2$] for the four
available gratings at seeing- or diffraction limited resolution.
When OH suppression was applied, the spectrum was resampled to a resolution
of R=400. In the case ``Extended source'', the integrated spectrum of the
pixels within the FWHM of the source (here set to be 10pix, 2\farcs5 for seeing,
0\farcs25 for AO-scale) was computed. See the text for further 
details. The given numbers are average values over the bands.}
\begin{center}
\renewcommand{\arraystretch}{1.6} \centering
\begin{tabular}{|c|c|c|c|c|c|c|c|}
\hline
\multicolumn{2}{|c|}{}  & \multicolumn{2}{c|}{Point Source} &  \multicolumn{2}{c|}{Extended ($\oslash$ 10 pix)} &  
\multicolumn{2}{c|}{per pixel}\\\cline{3-8}
\multicolumn{2}{|c|}{}  & No OH-sup. & OH-sup. & No OH-sup. & No OH-sup. & No OH-sup. & OH-sup.\\\hline\hline
 & Seeing & 20.4 & 22.7 & 21.0 & 23.3 & 18.4 & 20.7\\\cline{2-8}
\raisebox{2ex}[-2ex]{J}  & AO     & 22.2 & 24.3 & 17.4 & 19.6 & 14.8 & 17.0 \\\hline
 & Seeing & 19.6 & 22.0 & 20.3 & 22.5 & 17.8 & 20.0 \\\cline{2-8}
\raisebox{2ex}[-2ex]{H}  &  AO     & 21.7 & 23.7 & 17.0 & 19.0 & 14.4 & 16.4\\\hline
 & Seeing & 18.5 & 20.8 & 19.2 & 21.3 & 16.7 & 18.7\\\cline{2-8}
\raisebox{2ex}[-2ex]{K}  &  AO     & 21.0 & 22.9 & 16.3 & 18.2 & 13.7 & 15.6 \\\hline
 & Seeing & 19.5 & & 20.0 & & 17.5 & \\\cline{2-8}
\raisebox{2ex}[-2ex]{H+K} & AO     & 22.1 & & 17.4 & & 14.7 & \\\hline
\end{tabular}
\label{sensitivities}
\end{center}
\end{table}

\section{RESEARCH PROSPECTS}\label{research}

Now we will describe examples for the possible scientific targets
of interest and the observing modes that should be chosen for their
observations. Two of them are accompanied by simulated observations.

\subsection{Faint, extended objects}
For extended objects, the sensitivity of \spiffi\ is not
increased going to AO scales, because even though the sky background is reduced,
the source is distributed over a larger number of pixels accordingly.
In order to observe faint, extended objects, the larger
pixel size is required to maximize the sensitivity. 
This is true, no matter if the integrated
spectrum is taken or if the object is analyzed pixel-by-pixel.

One group of very obvious targets are high redshift galaxies, like
they are observed in the Hubble Deep Field (HDF) at a typical
spatial extent of a few arcsec. The determination of their
properties will have an impact on the theories of the formation and
evolution of galaxies. Their nature can
only be verified spectroscopically, and many of them
show a distorted spatial structure that underlines the benefit of 
spatially resolved spectroscopic information.
The advantage that the NIR provides for this type of objects is that
the optical diagnostic lines, like H$\alpha$ and $[$O III$]$, are
redshifted into the SINFONI observing windows for redshifts 
z$=0.5--2.7$ (H$\alpha$) and 1.0-3.9 ($[$O III$]$). 
H$\alpha$ is a tracer for star formation and not as prone to
extinction as Ly$\alpha$, while $[$O III$]$ serves as an excitation
diagnostic.

\begin{table}[ht]
\caption{Redshift ranges in which the lines in column 1 are observable in the
\sinfoni\ windows J, H and K.}

\begin{center}
\renewcommand{\arraystretch}{1.6} \centering
\begin{tabular}{|l|c|c|c|c|}
\hline
Line & $\lambda_0[\mu$m]& J & H & K \\\hline\hline
H$\alpha$ & 0.6563 & 0.5--1.1 & 1.5--1.8 & 2.0--2.7\\\hline
[OIII]5007 & 0.5007 & 1.0--1.8 & 2.0--2.7 & 3.0--3.9\\\hline
H$\beta$ & 0.4861 & 1.1--1.9 & 2.1--2.8 & 3.1--4.0\\\hline
[OIII]3727 & 0.3727 & 1.7--2.8 & 3.0--3.9 & 4.4--5.5\\\hline
Ly$\alpha$ & 0.1215 & 7.2--10 & 11--14 & 15--19\\\hline
\end{tabular}
\label{redshifts}
\end{center}
\end{table}

To identify and measure those lines, a low spectral resolution
is sufficient (e.g., R=400), which means that the observations are performed
at a high spectral resolution and after OH suppression are rebinned along
the spectral axis to result in a lower spectral resolution, thereby
increasing the SNR. 

\begin{figure*}%[hb]
\begin{center}
%\begin{tabular}{c}
\begin{minipage}[t]{8.2cm}
\hfill
\psfig{file=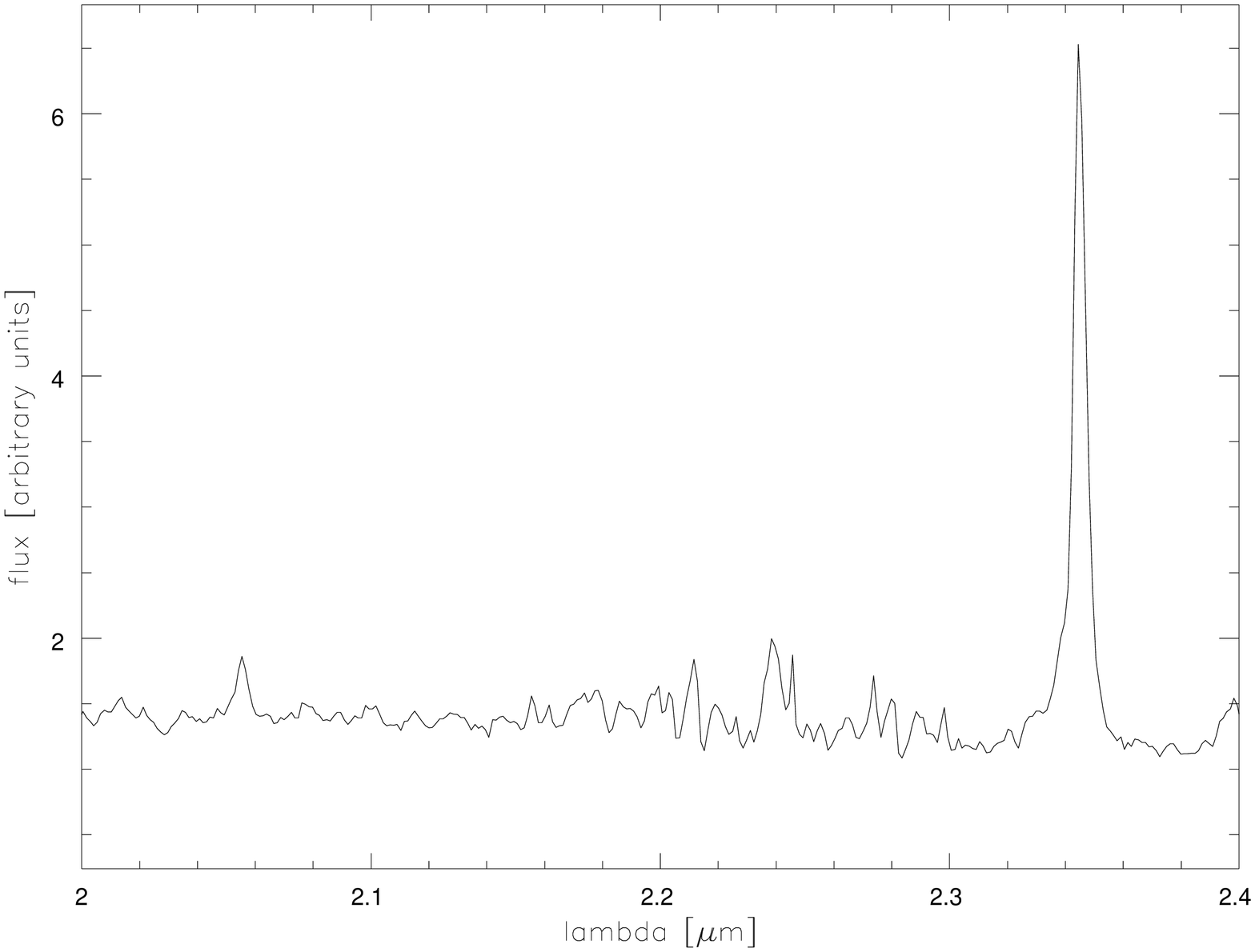,width=8.2cm,angle=-90}
\hfill
\end{minipage}
\vfill
\begin{minipage}[b]{8.2cm}
\psfig{file=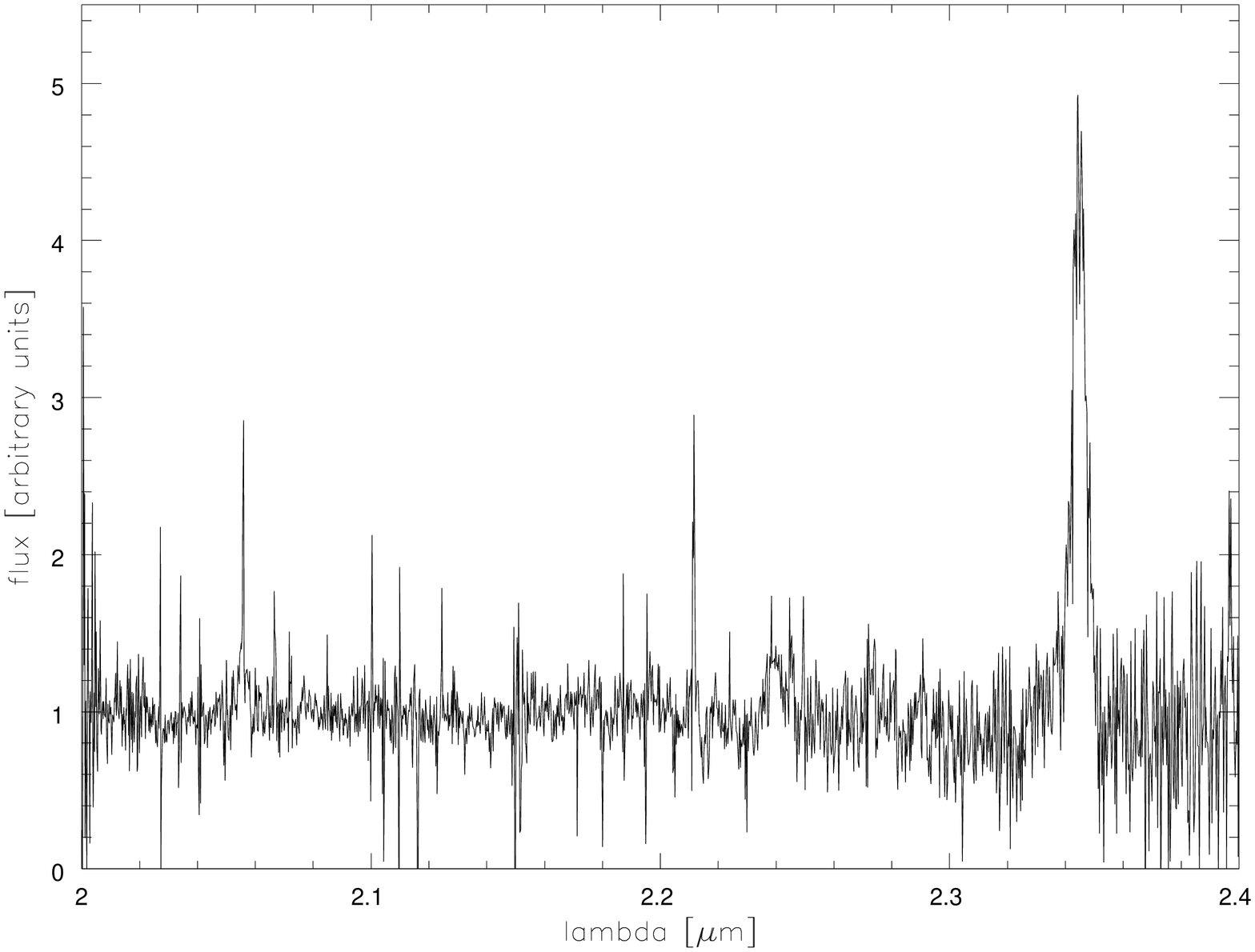,width=8.2cm,angle=-90}
\end{minipage}
\hfill
\begin{minipage}[b]{8.2cm}
\psfig{file=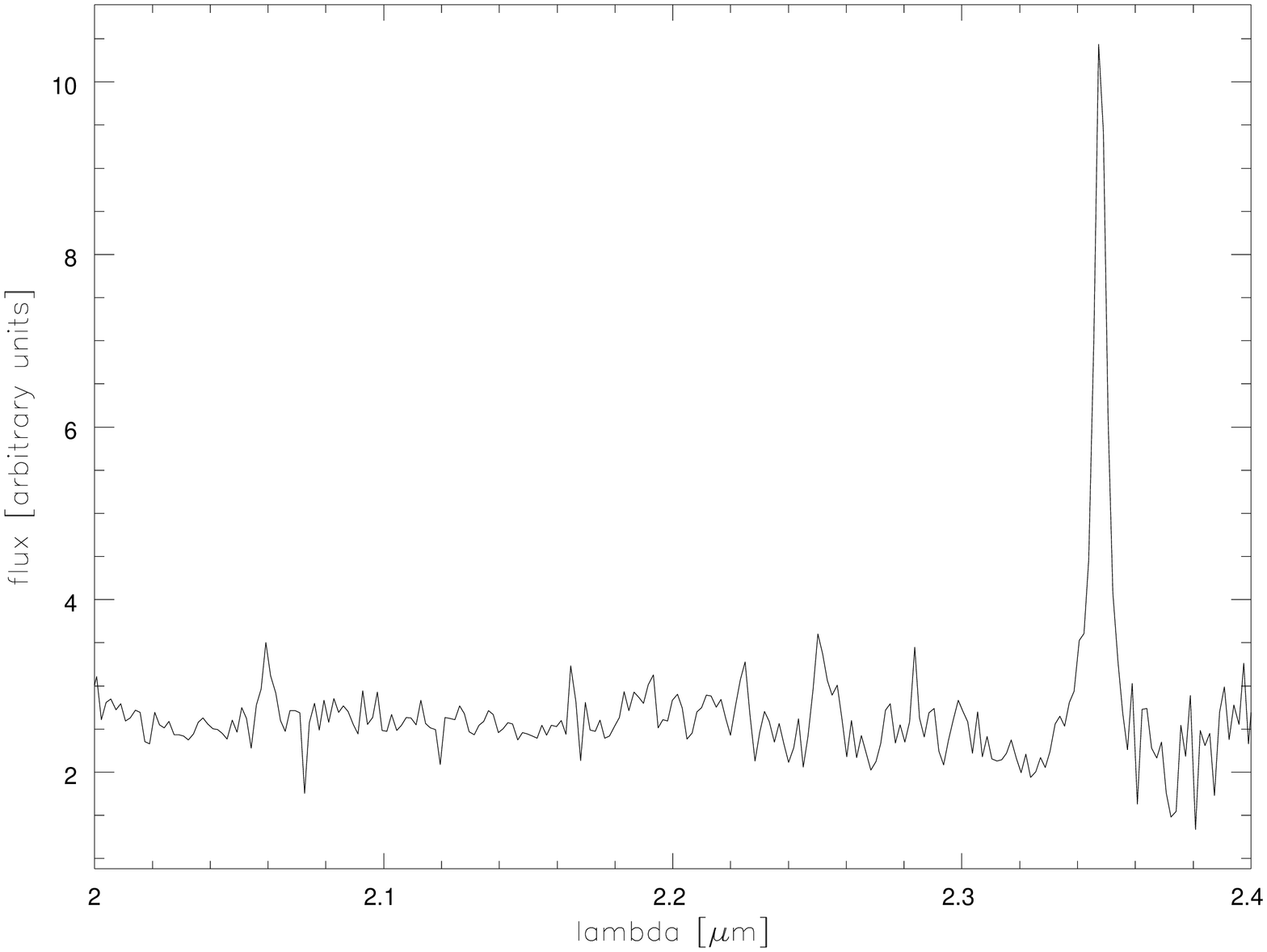,width=8.2cm,angle=-90}
\end{minipage}
%\end{tabular}
\end{center}
\caption{Simulation of \sinfoni\/ observations of an object with
a bright emission line (a ULIRG spectrum redshifted to  z=0.25, 
kindly provided by H. Dannerbauer). The input spectrum is displayed at the
upper  left. The output spectra
were achieved at one hour integration time, and show the spectra for
a single pixel. The surface brightness is 16 mag arcsec$^{-2}$. The lower left
spectrum has the original K-band resolution of R$\sim4500$, the lower 
right was resampled to R$\sim1300$, with software
OH suppression applied. The Pa$\alpha$ line
at $\sim2.35\mu$m can be identified in both spectra, providing information
about the star formation history of these objects.}
\label{ulirg_sim}
\end{figure*}

\begin{figure*}[ht]
\begin{center}
\begin{minipage}[t]{8.2cm}
\psfig{file=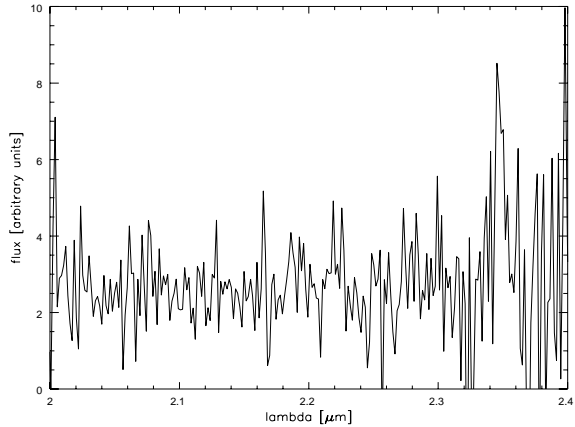,width=8.2cm,angle=-90}
\end{minipage}
\end{center}
\caption{Same as Figure \ref{ulirg_sim}, but now for a point source of
20 mag. Spectral resampling to R$\sim1300$ was applied. The line can just
be identified (S/N$\sim$5 in the line).}
\label{ulirg_sim2}
\end{figure*}

The spectra in Figures \ref{ulirg_sim} and \ref{ulirg_sim2}
show the simulated observations
of a spectrum with a strong emission line (initially an Ultra Luminous Infrared 
Galaxy (ULIRG) spectrum redshifted to 
z=0.25, where the Pa$\alpha$ line is shifted into the K-band). The simulated
performance is calculated for a single pixel in 1h integration time
for a source of 16 mag arcsec$^{-2}$ surface brightness. The top spectrum shows the
input spectrum, the result at the
initial resolution of R$\sim$4500 without OH-suppression is 
displayed in the lower left spectrum. The lower right spectrum was created by
 resampling (R$\sim$1300) with software OH suppression. 
For comparison, a spectrum derived at a point source magnitude of
20mag is plotted in Figure \ref{ulirg_sim2}. In all spectra, the Pa$\alpha$
line can be clearly identified and used as a diagnostic for star formation.

%\pagebreak

\subsection{Faint point sources}
In the case of point sources, the sensitivity is increased going to
the AO pixel scales, because the sky background is reduced and the
number of pixels covered by the source stays the same. 
So for maximum sensitivity, the AO pixel scale will be chosen.
In  the case of very faint targets, it is possible to convolve the
spectrum down to a lower resolution, if for example the detection
of emission lines is the goal.

Typical objects for this case are QSOs. Many questions concerning
their nature are still unresolved, like the existence and nature of their hosts
and the relative contributions of the quasars and their hosts to the
spectrum. The quasars also show strong emission lines which are redshifted
into the \sinfoni\ windows at various redshifts (see Table \ref{redshifts}).
They can provide clues about the nature of these objects.

\subsection{Bright extended objects}
The objects in this group are expected to be observed at high
spectral and spatial resolution. The targets are
for example interacting galaxies, the narrow line regions of AGN or
near-by ULIRGS as extragalactic targets, jets and disks around young
stars or planetary nebulae in our Galaxy. Those objects will profit
from the spatial coverage of \sinfoni, and their spectra can be studied
pixel-by-pixel.

One group of targets where already \sinfoni's precursor MPE 3D\cite{Weitzel96}
has been proven to deliver valuable data is the centres of external
galaxies. MPE 3D's spectral resolution of 2000 is sufficient to determine the
stellar dynamics in these targets, and it was shown that the three-dimensional
information is required for the determination of a central mass, because
only then can anisotropy effects be accounted for.\cite{Anders99}

%\begin{figure*}[hp]
%\begin{center}
%\begin{tabular}{c}
%\psfig{file=sinf_bh.ps,width=14cm}
%\end{tabular}
%\end{center}
%\caption{Simulated observations of a typical galactic nucleus. The
%parameters of the observations are given in the figure. The bottom panel
%shows the simulated spectrum of the galactic nucleus with the
%stellar CO bandhead features. Overlaid is 
%the fit of a stellar template which is used for the derivation of the
%velocity V and the dispersion $\sigma$. The top panels show those
%values as they are expected with and without the presence of a central
%dark mass. Those two models can be clearly distincted using the
%\sinfoni\/
% observations.}
%\label{bh_sim}
%\end{figure*}

\subsection{Bright point sources}
\begin{figure*}[ht]
%\begin{center}
%\begin{tabular}{c}
\begin{center}
\begin{minipage}[t]{8.3cm}
\hfill
\psfig{file=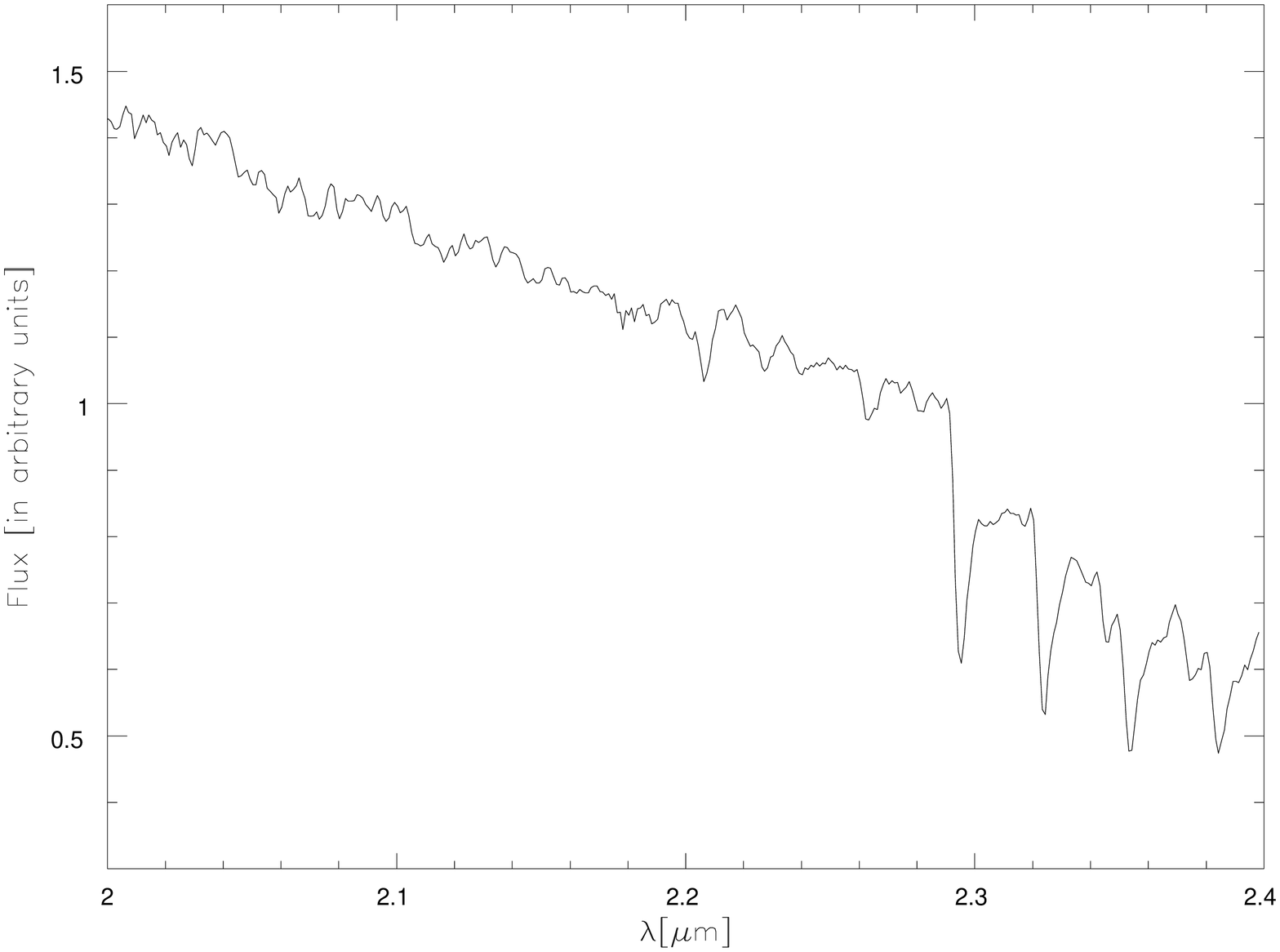,width=8.3cm,angle=-90}
\hfill
\end{minipage}
\end{center}
\vfill
\begin{minipage}[b]{8.3cm}
\psfig{file=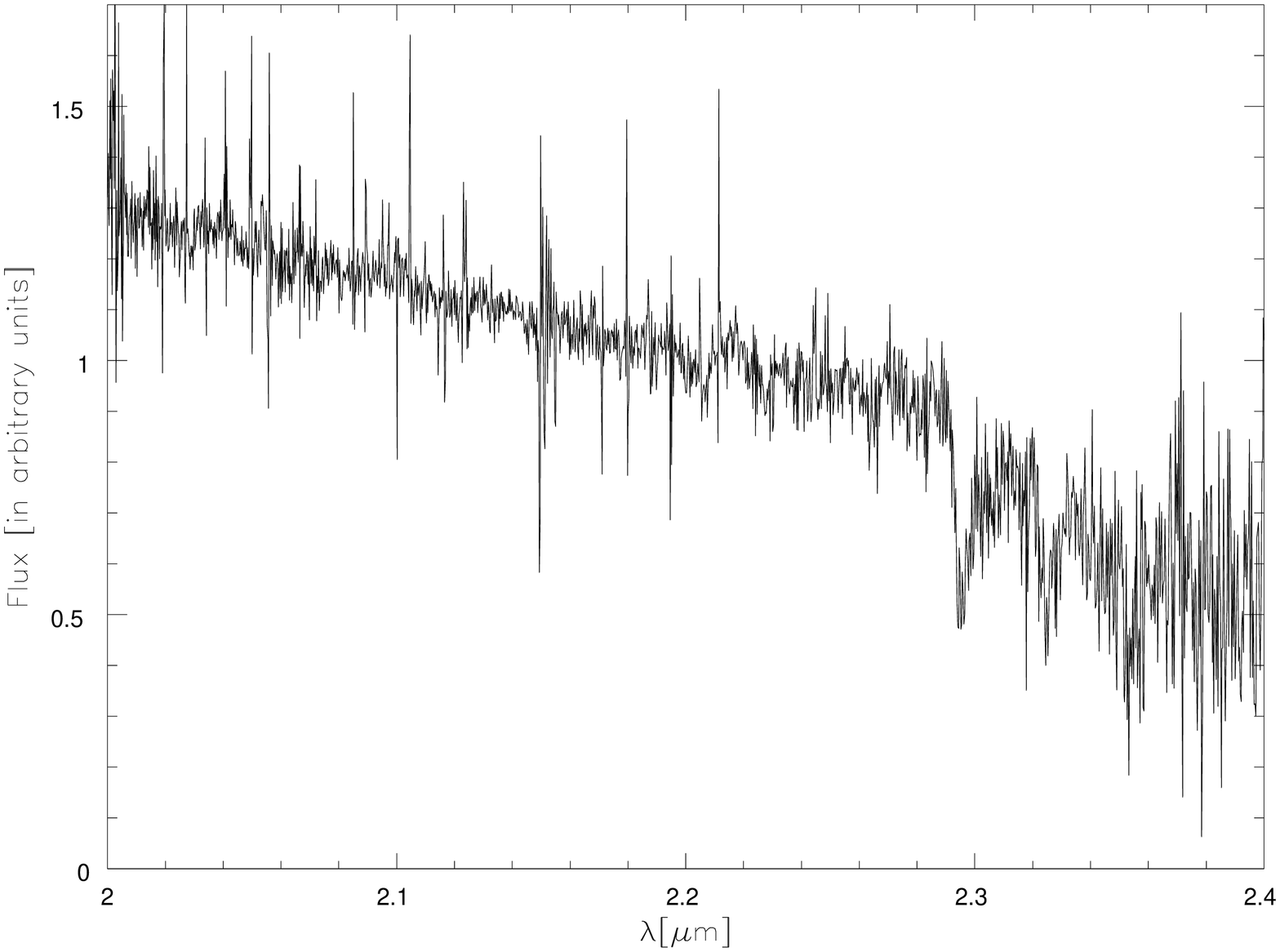,width=8.3cm,angle=-90}
\end{minipage}
\hfill
\begin{minipage}[b]{8.3cm}
\psfig{file=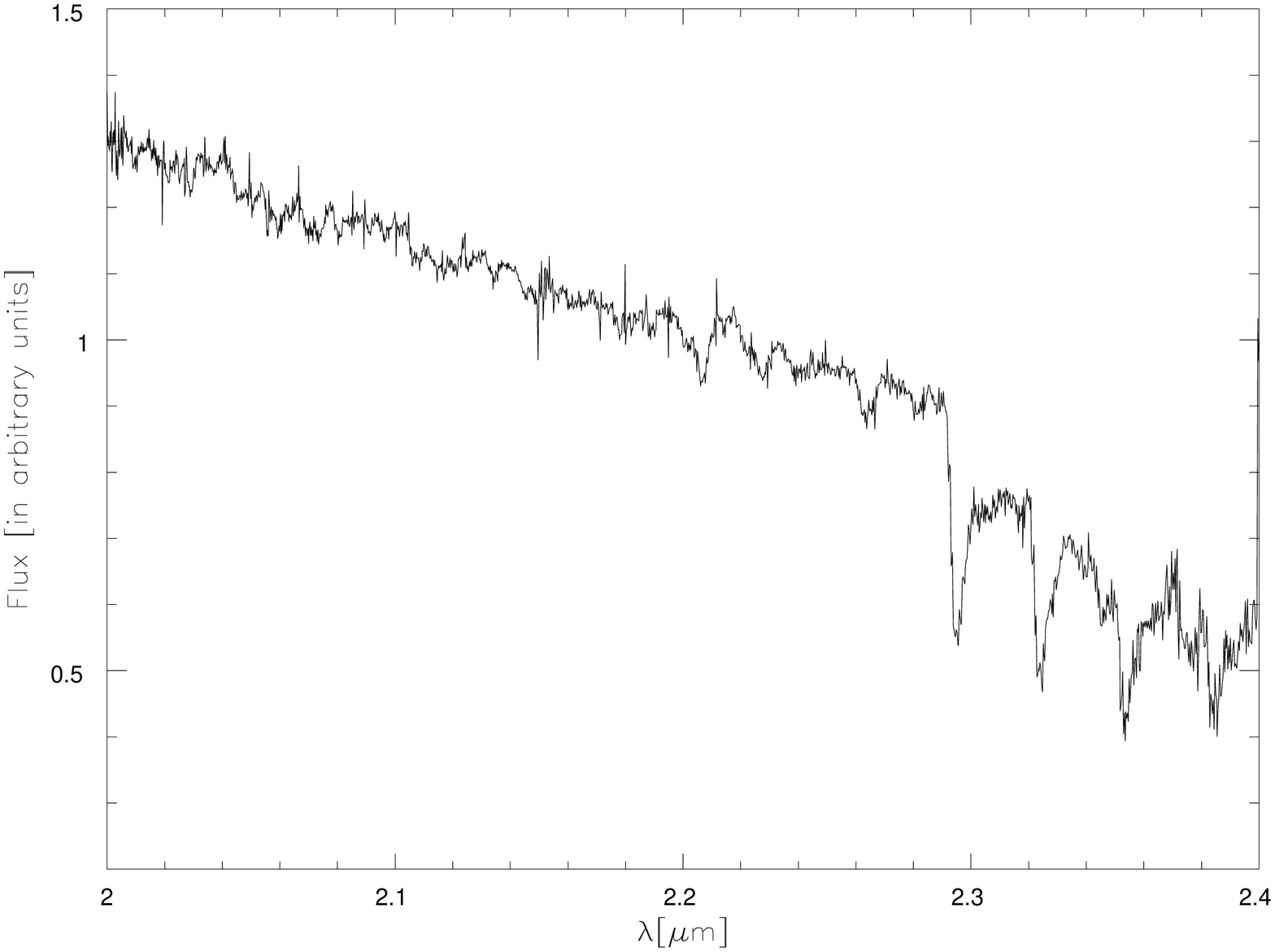,width=8.3cm,angle=-90}
\end{minipage}
%\end{tabular}
%\end{center}
\caption{Simulation of \sinfoni\/ observations of an M2Iab-Ib supergiant. 
The input spectrum is displayed at the top, the object is assumed to be
a point source of 17th mag (lower left) and 15th mag (lower right).
The integration time was one hour, no spectral rebinning was applied.
In the higher noise case, the presence of absorption can be verified,
in the low noise case the SNR ($\sim$50) is sufficient to derive a velocity 
dispersion}
\label{CO_sim}
\end{figure*}

For bright point sources, the possibility to use the AO scales
applies, as well, and the high spectral resolution can be exploited.
Possible targets are the cores of globular or young stellar clusters, close
binary star systems or star clusters in external galaxies.
In cases where the star clusters are resolved into single stars,
it is possible to determine their spectral type and thus to place them
on the Hertzsprung-Russell diagram. This will then allow conclusions on the stellar
content of these clusters and thereby the Initial Mass Function (IMF), which is still an open
question. 
The following simulation (Figure \ref{CO_sim}) 
shows a spectrum of an M supergiant (M2Iab-Ib).
This kind of spectrum is observed in clusters at an age in the 10Myr regime,
which are dominated by this type of stars. They can for example
be observed in interacting galaxies, or can be the central cluster in a galaxy.
If the velocity dispersion is high enough ($\sigma > \sim50$km/s)
in the cluster, it can be determined
at the spectral resolution of \spiffi. The input parameters for the simulation
were the standard parameters again, the object was assumed to be a point source
of 17th mag and 15th mag, respectively. 
The input spectrum is displayed at the top. No spectral rebinning
was applied. In the first case, the presence of CO absorption features can
clearly be verified, which can serve to age date the population.
In the second case, the SNR is high enough ($\sim$30) to determine a
broadening of the feature due to a velocity dispersion in the cluster.

%%%%%%%%%%%%%%%%%%%%%%%%%%%%%%%%%%%%%%%%%%%%%%%%%%%%%%%%%%%%%
%\acknowledgments     %>>>> equivalent to \section*{ACKNOWLEDGMENTS}       

%%%%%%%%%%%%%%%%%%%%%%%%%%%%%%%%%%%%%%%%%%%%%%%%%%%%%%%%%%%%%
%%%%% References %%%%%

\bibliography{report}   %>>>> bibliography data in report.bib
\bibliographystyle{spiebib}   %>>>> makes bibtex use spiebib.bst

\end{document}